\documentclass{elsart}
\usepackage{amssymb,bm,amsmath,graphicx,color,subfigure,url,lineno}
\usepackage{epstopdf}
\newcommand \be{\begin{equation}}
\newcommand \ee{\end{equation}}
\journal{Chaos, Solitons \& Fractals}%

\begin{document}

\runauthor{Zhou and Sornette} \markboth{A}{B}

\begin{frontmatter}

\title{Numerical investigations of discrete scale invariance in fractals and multifractal measures}

\author[ecust]{{Wei-Xing Zhou}},
\ead{wxzhou@ecust.edu.cn} %
\author[ETH]{{Didier Sornette}\corauthref{cor}}
\corauth[cor]{Corresponding author. Address: KPL F 38.2, Kreuzplatz
5, ETH Zurich, CH-8032 Zurich, Switzerland. Phone: +41 44 632 89 17,
Fax: +41 44 632 19 14.}
\ead{dsornette@ethz.ch}%
\ead[url]{http://www.er.ethz.ch/}%

\thanks[grants]{This work was partially supported by the NSFC
(Grant 70501011) and the Fok Ying Tong Education Foundation (Grant
101086).}

\address[ecust]{School of Business, School of Science, and Research
Center of Systems Engineering\\East China University of Science and
Technology, Shanghai 200237, China}
\address[ETH]{D-MTEC, ETH Zurich, CH-8032 Zurich, Switzerland}

\begin{abstract}
Fractals and multifractals and their associated scaling laws provide
a quantification of the complexity of a variety of scale invariant
complex systems.  Here, we focus on lattice multifractals which
exhibit complex exponents associated with observable
log-periodicity. We perform detailed numerical analyses of lattice
multifractals and explain the origin of three different scaling
regions found in the moments. A novel numerical approach is proposed
to extract the log-frequencies. In the non-lattice case, there is no
visible log-periodicity, {\em{i.e.}}, no preferred scaling ratio
since the set of complex exponents spread irregularly within the
complex plane. A non-lattice multifractal can be approximated by a
sequence of lattice multifractals so that the sets of complex
exponents of the lattice sequence converge to the set of complex
exponents of the non-lattice one. An algorithm for the construction
of the lattice sequence is proposed explicitly.
\end{abstract}



\end{frontmatter}

\section{Introduction}
\label{s1:intro}

The word fractal was coined by Mandelbrot to describe sets, which
consist of parts similar to the whole and which can be described by
a fractal dimension \cite{Mandelbrot-1983}, which is in most
cases a fractional number. Meanwhile, multifractals were introduced
to characterize the statistics of the strength of singularity of
self-similar measures carried on general geometric supports \cite{Mandelbrot-1974-JFM,Grassberger-1983-PLA,Hentschel-Procaccia-1983-PD,Frisch-Parisi-1985,Halsey-Jensen-Kadanoff-Procaccia-Shraiman-1986-PRA}.
Fractals and multifractals have been extensively introduced in the
analysis of a wide class of phenomena in physics, geophysics,
chemical engineering, turbulence, growth phenomena, and so on. At
the same time, new ingredients were added in the theory of
multifractals. Based on the discovery of anomalies in DLA (Diffusion Limited
Aggregation), the phenomenon of phase transition in the multifractal spectrum was
found and studied
\cite{Lee-Stanley-1988-PRL,Blumenfeld-Aharrony-1989-PRL,Schwarzer-Lee-Bunde-Halvin-Roman-Stanley-1990-PRL}.
This led to the mathematical framework of the so-called left-sided
multifractals
\cite{Mandelbrot-1990b-PA,Mandelbrot-Evertsz-Hayakawa-1990-PRA,Riedi-1995-JMAA,Riedi-Mandelbrot-1995-AAM}.
The vector-valued multifractals were introduced to formulate
nonconservative measures such as the velocity field of turbulence
\cite{Falconer-ONeil-1996-PRSA}. Continuous multifractals
\cite{Zhou-Yu-2001-PA,Zhou-Liu-Yu-2001-Fractals,Huillet-Porzio-2001-Fractals}
were adopted to quantify the droplets breakup in turbulent jet flows
\cite{Zhou-Yu-2001-PRE}, which generalize the multipliers from
discrete to continuous.

Another important concept is the negative dimension or latent
dimension
\cite{Mandelbrot-1989-PAG,Mandelbrot-1990a-PA,Mandelbrot-1991-PRSA}.
The physical implication of negative dimensions was discussed
firstly in \cite{Cates-Witten-1987-PRA}. In practice, most of the
physical processes are random, which leads to the phenomenon of
sample-to-sample fluctuations of the multifractal function
$f(\alpha)$ by an amount greater than what the error bars on any
single sample would indicate. In general, negative dimensions arise
in such random multifractals
\cite{Chhabra-Sreenivasan-1991-PRA,Chhabra-Sreenivasan-1992-PRL}.
The randomness of multifractals is found at least in two cases
where negative dimensions may emerge
\cite{Chhabra-Sreenivasan-1991-PRA,Chhabra-Sreenivasan-1992-PRL}.
First, one may obtain a multifractal constructed by a multiplicative
cascade that is inherently probabilistic and the negative
dimensions, if they exist, describe the rarely occurring events.
Second, one may have to investigate the experiment from a
probabilistic viewpoint. For instance, one-dimensional cuts of a
deterministic measure carried out using a deterministic Sierpinski
sponge will inevitably introduce randomness and may be regarded as
random samples of a population. It is worth remarking that
randomness in multinomial measures does not imply the existence of
negative dimensions \cite{Zhou-Yu-2002}. Actually, the multifractal
slice theorem \cite{Olsen-1999-HMJ,Olsen-2000-PP} presents a
mathematical interpretation of negative dimensions and is the basis
of experimental measurement of lower dimensional cuts when the
measurement of the whole set is difficult to perform.
The multifractal slice theorem makes
rigorous some aspects of Mandelbrot's intuition that negative
dimensions may be explained geometrically by considering cuts of
higher dimensional multifractals.

In the context of fractals and multifractals, scale invariance is
the most important concept of symmetry to characterize fractals or
multifractal measures. In its simplest form, scale invariance can be
expressed by the following expression:
\begin{equation}
 {\mathcal{O}}(r) = \ell {\mathcal{O}}(\lambda r), \label{Eq:O}
\end{equation}
where $\mathcal{O}$ is the observable and $\lambda$ is the
magnification factor or scaling ratio. In the common sense,
$\lambda$ is a continuous parameter resulting in the so-called
continuous scale invariance (CSI). In other words, CSI is
associated with the fact that expression (\ref{Eq:O}) holds
for arbitrary values of $\lambda$ chosen in the set of positive
real numbers. However, CSI is not always true
when investigating fractals and multifractals. The general solution
to Eq.~(\ref{Eq:O}) is not just a power law as CSI would imply but
take the form
\begin{equation}
{\mathcal{O}}(r) = r^{-D} \psi(r), \label{Eq:OSol}
\end{equation}
where $D=\ln  \ell / \ln \lambda $ is the fractal dimension and $\psi(\lambda r) = \psi(r)$
is a priori an arbitrary log-periodic function with period $\ln\lambda$
\cite{Bessis-Geronimo-Moussa-1983-JPL,Badii-Politi-1984-PLA,Smith-Fournier-Spiegel-1986-PLA,Bessis-Fournier-Servizi-Turchetti-Vaienti-1987-PRA,Bessis-Servizi-Turchetti-Vaienti-1987-PLA,Fournier-Turchetti-Vaienti-1989-PLA}
(the term ``log-periodic'' means that the function $\psi$ is
periodic in the variable $\ln r$ with period $\ln \lambda$).
Actually, the recently proposed discrete scale invariance (DSI) and
its associated complex dimensions are an elaboration of the concept
of scale invariance in which that system is scale invariant only
under powers of specific values of the magnification factor (see
\cite{Sornette-1998-PR} for a review). DSI with the signature of
log-periodic oscillations decorating power laws have been reported in
several practical applications, such as in material rupture
\cite{Johansen-Sornette-1998-IJMPC,Johansen-Sornette-2000b-EPJB,Zhou-Sornette-2002-PRE},
earthquake precursors
\cite{Johansen-Saleur-Sornette-2000-EPJB,Huang-Saleur-Sornette-2000-JGR},
DLA
\cite{Sornette-Johansen-Arneodo-Muzy-Saleur-1996-PRL,Johansen-Sornette-1998-IJMPC},
turbulence
\cite{Novikov-1990-PFA,Johansen-Sornette-Hansen-2000-PD,Zhou-Sornette-2002-PD},
and economics \cite{Sornette-2003}. In addition, there are also
reports in the theoretical aspects of fractals
\cite{Bessis-Geronimo-Moussa-1983-JPL,Badii-Politi-1984-PLA,Smith-Fournier-Spiegel-1986-PLA,Bessis-Fournier-Servizi-Turchetti-Vaienti-1987-PRA,Bessis-Servizi-Turchetti-Vaienti-1987-PLA,Fournier-Turchetti-Vaienti-1989-PLA}
and multifractals \cite{Zaslavsky-2000-PA}. Notice the evolution of
the concept of dimension from integer to fractional to negative and
finally to complex numbers, each time with precise physical meaning.

However, Eq.~(\ref{Eq:O}) is only a special (lattice) case of DSI (when
$\lambda$ takes discrete values)
\cite{Lapitus-Frankenhuysen-2000}. In this work, we shall review the
general framework of DSI in strictly self-similar fractals and
multifractal measures in Appendix~\ref{s1:framework} with extension
to joint multifractal measures based on the self-similarity of
moments. The structure of the complex exponents will also be
discussed for both lattice and non-lattice cases.

The rest of this
paper is organized as follows. Extensive numerical simulations are
carried out on lattice multifractal measures to investigate
different scaling regimes stemming from finite size effects in
Sec.~\ref{s1:NumSim}. We propose a novel approach for the extraction
of log-periodicity in the lattice case in Sec.~\ref{s1:DSItest}.
Sec.~\ref{s1:concl} concludes. The Appendix summarizes the
general framework of discrete scale
invariance in fractals and multifractals, which is used throughout the paper.

\section{Numerical simulations on lattice multifractals}
\label{s1:NumSim}

In the absence of analytic results, this section presents
computer simulations to explore the
properties of lattice multifractal measures.  The properties of the
log-periodic function $\psi(r)$ are investigated in details. This task remains
difficult for lattice multifractals
with more than three partitions in each generation. With the
increase of partition number $n$, the iteration number and hence
the computational time needed in order to obtain
detectable log-periodic structure with relatively low noise level
increases exponentially. We thus have
to confine our study to small $n$ values. In
Sec.~\ref{s2:Finite}, we study binomial measures with
fixed amplification factors $\lambda_1 = (\sqrt 5 +1)/2$ and
$\lambda_2 = (\sqrt 5 +3)/2$ and varying measure multipliers
$m_1=m$ and $m_2=1-m$ to demonstrate three scaling regimes: the
fractal regime, the crossover regime and the Euclidean regime, the latter
resulting from finite size effects. Note that the
geometric support is not fractal. Then, we study the properties of
the Euclidean regime in Sec.~\ref{s2:ResIII}, allowing us to gain
insight into the structure of the noise stemming from the incomplete
construction of the multifractal measure and of the partitioning. In
Sec.~\ref{s2:ResI}, we study the scaling properties of the
amplitude of log-periodic oscillations decorating the scaling of
moments.

\subsection{\label{s2:Finite}Finite size effect}

We performed 13 iterations to construct a lattice binomial measure with
measure multipliers $m=0.2$ and $1-m=0.8$
(see \cite{Mandelbrot-1989a,Riedi-Mandelbrot-1995-AAM,Halsey-Jensen-Kadanoff-Procaccia-Shraiman-1986-PRA} for the definition and construction of binomial and
multinomial measures). Then the moments $\Gamma(r,q)$ for different orders
$q$ were calculated based on the box-counting method with sizes $r$
evenly distributed in logarithmic scale in $[e^{-9},e^{-1}]$ (see the Appendix
for definitions). The log-log
plot of $\Gamma(r,q)$ against $r$ for $q=1$ recovers
$\Gamma(r,q=1) = 1$ (as it should from the normalization
of the measure and the definition (\ref{Eq:MfMom}) of the moments)
with minor fluctuations of order $10^{-15}$
stemming from the precision of the computation. The log-log plot of $\Gamma(r,q)$ versus
$r$ for $q=0$ is a perfect straight line with slope $-D_0=-1$. Recall that, if
the geometric support is fractal, the slope should be $-D_f$. The
log-log plot of $\Gamma(r,q)$ against $r$ for $q \in (0,1) \cup
(1,\infty)$ looks like a strip with slope $\tau(q)$ which is
analogous to the situation presented in Fig.~\ref{Fig10}. We postpone the discussion of this
case till Sec.~\ref{s1:DSItest}.

\begin{figure}
\begin{center}
\includegraphics[width=8cm]{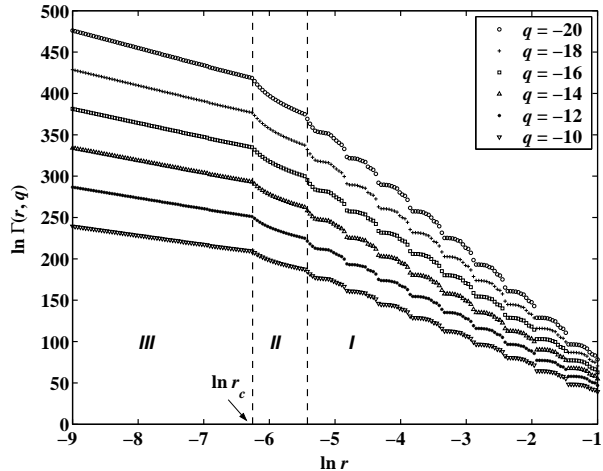}
\end{center}
\caption{\label{Fig1} The dependence of moments $\Gamma(r,q)$ with
respect to box size $r$ in a log-log plot for $q<0$. The
multifractal is a lattice binomial measure with scale multipliers
$\lambda_1 = \sqrt{\lambda_2} = (\sqrt{5}+1)/2$ and measure
multipliers $m_1 = 1-m_2 = 0.2$. We performed 13 iterations to
construct the multifractal. There are three regimes clearly visible
separated by two vertical dashed lines, where ${\it{I}}$ is the
fractal regime, ${\it{III}}$ is the Euclidean regimes and
${\it{II}}$ is the crossover regime. The left vertical line is
located precisely at $r_c = r_1^{13}$ (see text for details).}
\end{figure}

Fig.~\ref{Fig1} shows the dependence of the moments $\Gamma(r,q)$
with respect to box size $r$ in a log-log plot for $q<0$. There
are three regimes clearly visible in the figure separated by two
vertical dashed lines. Log-periodic oscillations decorate the
scaling law in regime ${\it{I}}$. It is easy to calculate
numerically the corresponding log-frequency $f_0 \approx
1/\ln\lambda = 2.0781$ by employing a spectral analysis on the
residuals defined as $\psi-\langle \psi \rangle$ or on the local
derivatives defined in Eq.~(\ref{Eq:MfDeftaur}) in the Appendix. Regime $\it{III}$
corresponds to perfect straight lines with different slopes.
Regime $\it{II}$ shows the crossover from large box sizes in
regime ${\it{I}}$ to small ones in regime ${\it{III}}$. We shall
clarify in the sequel that ${\it{I}}$ is the fractal regime,
${\it{III}}$ is the Euclidean regime and ${\it{II}}$ is the
crossover regime. We shall see that this phenomena is due to the
finite size effect.

It is well known that, although the mathematical model for the
construction of a fractal or multifractal measure contains an
infinite number of steps of the iterative construction process, the
notion that they are (statistically) self-similar can only apply
between certain cutoffs in practice \cite{Mandelbrot-1983}. In other
words, what we are able to deal with in nature, laboratory and
computer simulations are in fact {\it pre-fractals} and {\it
pre-multifractals}. We nevertheless use the term ``fractal'' and
``multifractal'' without the prefix ``pre'', following the custom in the
literature.

Let $r_c$ denote the crossover scale (close to the inner cutoff) of
a multifractal with measure $\mu_i=\mu(r_i)$ distributed uniformly
on a segment ${\mathcal{F}}_i$ of scale $r_i$, where $i = 1, \cdots,
n$. It is clear that \be r_c = \left(\max\{r_i\}\right)^b,
\label{Eq:rc} \ee where $b$ is the iteration number of the
construction. This $r_c$ is the largest segment among the $n^b$
segments with uniform measure. Choosing to count boxes with $r <
r_c$, we have approximately that $\Gamma(F; r,q) \approx
\sum_{i=1}^n \Gamma({\mathcal{F}}_i; r,q)$. There are about $n_i =
r_i/r$ boxes covering segment ${\mathcal{F}}_i$, each box possessing
the measure $\mu_i/n_i$. Thus, we have $\Gamma({\mathcal{F}}_i; r,q)
= n_i (\mu_i/n_i)^q$. It follows immediately that
\begin{equation}
\Gamma(F; r,q) \approx \sum_{i=1}^n \mu_ir_i^{1-q} r^{q-1}.
\label{Eq:FSE}
\end{equation}
where $\sum_{i=1}^n \mu_ir_i^{1-q}$ is independent of the box size
$r$. We obtain that $\tau(q) = q-1$ from the definition
(\ref{Eq:Mftau}) and $D_q = 1$, corresponding to the Euclidian regime ${\it{III}}$.
Concerning the general situation in a $d$-dimensional space,
we have $\tau(q) = (q-1)d$ and $D_q = d$ in
regime ${\it{III}}$. This result just states that a pre-fractal is Euclidean if we
investigate it at a very high resolution.

We have observed in Fig.~\ref{Fig1} that $r_c$ is precisely the
right boundary of the Euclidean regime $\it{III}$. The inner
cutoff is obviously larger than $r_c$, since a box with size
greater than $r_c$ doesn't guarantee the absence of a possible
destruction of the log-periodic structure close to $r_c$. Indeed,
such a box will inevitably split a segment into two parts which
introduces errors. However, the inner cutoff can not be determined
rigorously but only numerically.

When changing the value of $m$, we can still observe four
different kinds of moments $\Gamma(r,q)$ associated
with different $q$'s: (1)
$q < 0$, (2) $q = 0$, (3) $q = 1$ and (4)
$q\in(0,1)\cup(1,\infty)$. It is apparent that the plots of
the moments as a function of $r$ for (2) and (3) are similar for all
lattice and
non-lattice multinomial measures, and thus trivial. For the present
binomial measure, when $m$ is large (close to 1), case (1)
exhibits a strip while (2) shows three regimes. The results for
small $m$ on the three regimes also apply by symmetry $m \to 1-m$ to large $m \to 1$. When
$m=(\sqrt{5}-1)/2$, the measure degenerates to a mono-fractal, where
the measure is uniformly distributed over the support with density
$1$.

\subsection{\label{s2:ResIII}Residuals of the moments in the
Euclidean regime ${\it{III}}$}

\begin{figure}
\begin{center}
\includegraphics[width=8cm]{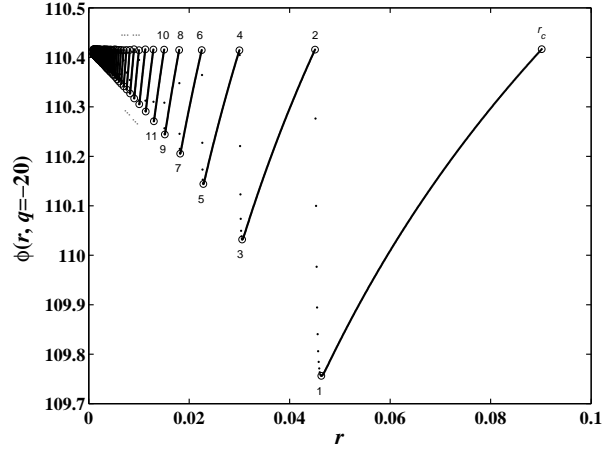}
\end{center}
\caption{\label{Fig2} The dependence of the residual $\phi(r,q=-20)$
defined by expression (\ref{Eq:phi})
with respect to $r$. The even numbers are used to label the points
of local maxima, while the odd numbers denote points of local
minima. The point $0$ is precisely at $r=r_c$. It is visible that
$\phi(r_{2n},q)$ is constant for a given $q$.}
\end{figure}

Although regime ${\it{III}}$ is trivial in the sense that $\tau(q)
= (q-1)d$, there are still fine structures stemming from those
boxes covering two segments and the rightmost box as well. Let us
define residuals in regime ${\it{III}}$ as
\be
\phi(r,q) =
\ln\Gamma(r,q) - (q-1)\ln r.
\label{Eq:phi}
\ee
 For different
$q<0$, the residuals $\phi(r,q)$ have similar shapes. Figure
\ref{Fig2} plots the residual for $q=-20$ with respect to $r$. We
performed $b=5$ iterations to construct the binomial measure. We
see accelerating oscillations with decaying amplitudes towards
$r=0$. The maximal box size is $r_c = 0.0902$. We labelled this
point as $0$. The consecutive points of local minima and local
maxima are labelled as $1, 2, \cdots$ in turn as shown in
Fig.~\ref{Fig2}. The coordinates of these points are denoted
accordingly $(r_i, \phi(r_i,q))$ with $i = 0,1,2,\cdots$. The
$n$-th oscillation is from point $2n-2$ through point $2n-1$ to
point $2n$. One can see that the oscillations are similar to each
other. We find that \be \phi(r_{2n},q) \approx \phi(r_c,q).
\label{Eq:phi2n} \ee Note that $r_0 = r_c$. The function
$\phi(r_c,q)$ is found to be linear with respect to $q$ with
negative slope.

\begin{figure}
\begin{center}
\includegraphics[width=8cm]{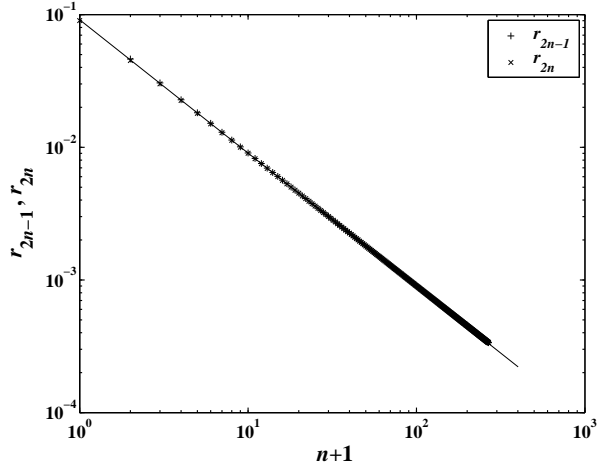}
\end{center}
\caption{\label{Fig3}The power law relationship of $r_{2n}$ (with $n
\in \overline{{\mathcal{Z}}^-}$) and $r_{2n-1}$ (with $n \in
{\mathcal{Z}}^+$) with respect to $n+1$. The scaling exponents are both
$-1$.}
\end{figure}

Figure \ref{Fig3} draws the box sizes $r_{2n-1}$ and $r_{2n}$ versus
$n+1$ in a log-log plot. We find the following power laws: \be
r_{2n-1} = 2r_1(n+1)^{-1} \label{Eq:r2n1} \ee for $n \in
{\mathcal{Z}}^+$ and \be r_{2n} = r_c(n+1)^{-1} \label{Eq:r2n} \ee
for $n \in \overline{{\mathcal{Z}}^-}$. It is important to note that
$r_{2n-1}$ and $r_{2n}$ are independent of $q$ for given binomial
measures. For different iterations $b$, $r_c$ and $r_1$ are not
constant anymore. We can determine $r_{2n}$ analytically by
combining Eqs.~(\ref{Eq:rc}) and (\ref{Eq:r2n}), while $r_1$ should
be calculated numerically for different $b$. Let $(\delta r)_n =
r_{2n}-r_{2n-2}$, then $(\delta r)_n = 1/n(n+1)$. When $n\to
\infty$, it follows that $(\delta r)_{n+1}/(\delta r)_n \to 1$,
which means that the oscillations are approximately periodic locally
for large $n$. This is not surprised since $(\delta r)_n$ decays
more and more slowly when $r \to 0$ following a $1/n^2$ law.

\begin{figure}
\begin{center}
\includegraphics[width=8cm]{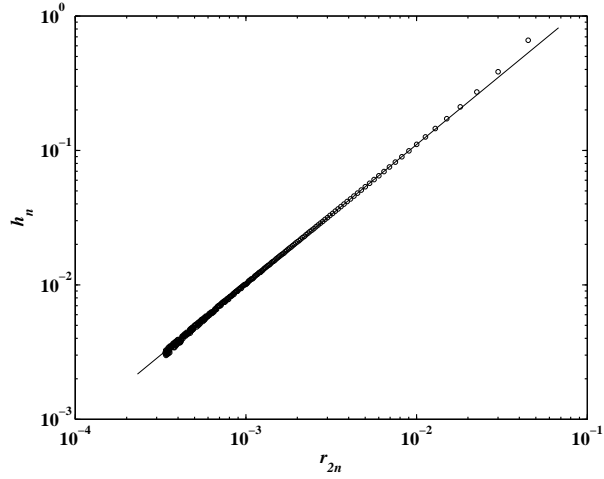}
\end{center}
\caption{\label{Fig4} The power law dependence of the amplitude
$h_n(q)$ with $q=-20$ versus $r_{2n}$. The scaling exponent is
$\alpha=1.0569$.}
\end{figure}

Define the amplitude of the $n$-th oscillation as the distance of
point $(r_{2n-1}, \phi(r_{2n-1},q))$ to the line $\phi(r,q) =
\langle\phi(r_{2n},q)\rangle_n$: \be h_n(q) = \phi(r_{2n-1},q) -
\langle\phi(r_{2n},q)\rangle_n, \label{Eq:hnq} \ee where
$\langle\rangle_n$ is an average over different oscillations $n$.
Figure \ref{Fig4} shows the dependence of the amplitude $h_n(q)$
for $q=-20$ with respect to $r_{2n}$. Again, we observe a
remarkable power law relationship: \be h_n(q) = c(q)\cdot
r_{2n}^\alpha,~~ \alpha = 1.0569, \label{Eq:hnqr} \ee where \be
c(q) = h_1(q)(r_c/2)^{-\alpha} \label{Eq:cq} \ee is a function of
$q$.

\begin{figure}
\begin{center}
\includegraphics[width=8cm]{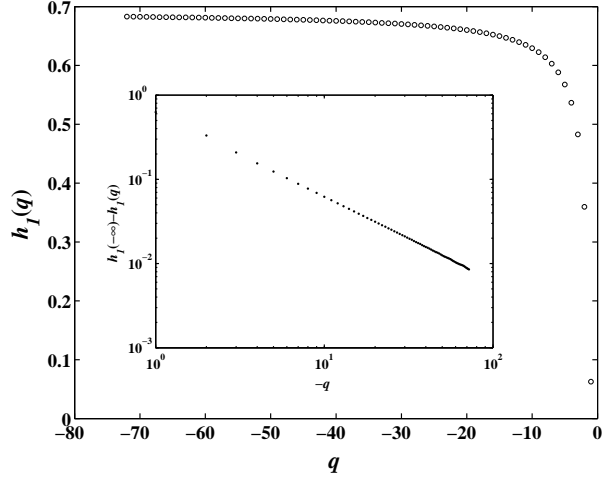}
\end{center}
\caption{\label{Fig5}The dependence of $h_1(q)$ with respect to $q$.
There is a limit for $h_1(q)$ when $q$ tends to $-\infty$. The inset
shows the power law dependence of $h_1(-\infty)-h_1(q)$ versus $-q$.
The scaling exponent is $\beta=1.0013$.}
\end{figure}

We also investigated the dependence of $h_1(q)$ with respect to
$q$, as shown in Fig.~\ref{Fig5}. It seems that $h_1(q)$
approaches a constant when $q$ tends to $-\infty$. The points
($\circ$) in Fig.~\ref{Fig5} can be represented by the following
law: \be h_1(q) = h_1(-\infty) - C/|q|^\beta, \label{Eq:h1q} \ee
where the values $h_1(-\infty) = 0.6915$, $\beta = 1.0013$, and $C
= 0.6293$ are obtained from a nonlinear regression procedure. The
inset verifies the power law dependence of $h_1(-\infty)-h_1(q)$
as a function of $-q$. We also find that $h_1(-\infty) = 0.6915$
is universal for different $b$ and $m$, while $\beta$ and $C$ vary
for different $b$ and $m$. In addition, the amplitude $h_1(q)$
increases with $|q|$.

\subsection{\label{s2:ResI}Residuals of the moments in the
fractal regime {\it{I}}}

\begin{figure}
\begin{center}
\includegraphics[width=8cm]{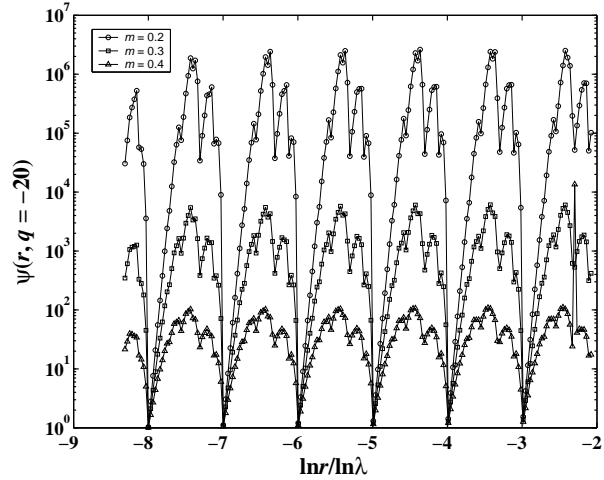}
\end{center}
\caption{\label{Fig6} The dependence of $\psi(r,q)$ for $q=-20$ with
respect to $\ln r/\ln\lambda$ for different measure multipliers $m$. The periodicity is
clearly visible with period $\ln\lambda$ in the $\ln r$ variable. The
amplitude of the oscillations decreases with increasing $m$ for
fixed $q$. The local minima are located precisely at $r =
\lambda^{-i}$ with $i \in \overline{{\mathcal{Z}}^-}$.}
\end{figure}

We now study the log-periodic function $\psi(r,q)$ and the
corresponding residuals $\ln \psi(r, q) = \ln\Gamma(r,q) -
\tau(q)\ln{r}$ of moments in regime ${\it{I}}$ defined in
Eq.~(\ref{Eq:MfSol}) of the Appendix. Denoting $\lambda_1 = \lambda$, we find
$\lambda_2 = \lambda^2$. From (\ref{Eq:CE}), we have $m^q
\lambda^{\tau(q)} + (1-m)^q \lambda^{2\tau(q)} =1$ which is a
solvable univariate second-order equation with respect to
$\lambda^{\tau(q)}$. The explicit expression of the exponent
$\tau(q)$ follows: \be \tau(q) =\frac
{\ln2-\ln\left[m^q+\sqrt{m^{2q}+4(1-m)^q}\right]}{\ln\lambda},
\label{Eq:tau} \ee where $\lambda = (\sqrt{5}+1)/2$ is the
preferred scaling ratio. The behavior of $\psi$ is simpler, since
it is strictly periodic in $\ln r$. Thus, given $m$ and $q$, the
log-periodic oscillations have constant amplitude as a function of $r$.
Figure \ref{Fig6} shows the dependence of $\psi(r,q)$ for $q=-20$
with respect to $\ln r/\ln\lambda$ for different measure multipliers $m$.

The three lines in Fig.~\ref{Fig6} are periodic in $\ln
r/\ln\lambda$ with period $1$. The amplitudes of the oscillations
decrease with increasing $m$ for fixed $q$. The local minima are
located precisely at $r = \lambda^{-i}$ with $i \in
\overline{{\mathcal{Z}}^-}$. The abscissa of the local maxima are
different for different $m$, while those for same $m$ but different
$q$ overlap. For large $r$, the noise increases with $m$. For small
$r$, the oscillations are gradually spoiled to the right of the
crossover regime $\it{II}$, which were not shown in Fig.~\ref{Fig6}.
Hence, it is better to investigate the oscillations in the middle of
the fractal regime ${\it{I}}$ to obtain a quantitative description
of the log-periodic oscillations of $\psi(r,q)$.

\begin{figure}
\begin{center}
\includegraphics[width=8cm]{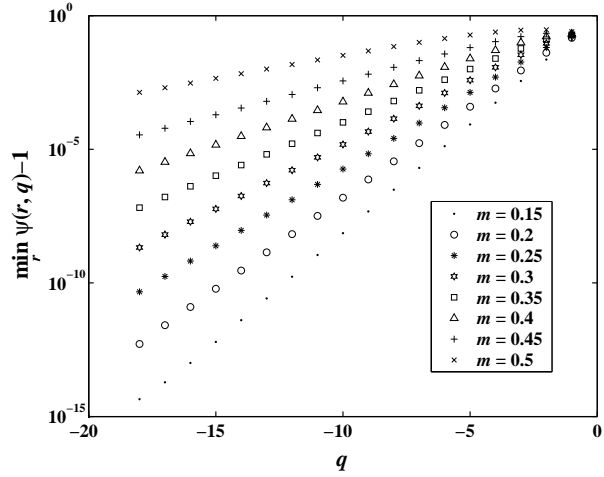}
\end{center}
\caption{\label{Fig7} The exponential approach of $\min_r\psi(r,q)$
to $1$ when $|q|$ increases for different $m$ is shown. The slopes
of the straight lines are approximately evenly spaced for small
$m$.}
\end{figure}

As shown in Fig.~\ref{Fig6}, the local minima ${\min}_{r}\psi(r,q)$
are close to $1$ at $r=\lambda^{-i}$ with $i \in
\overline{{\mathcal{Z}}^-}$. Actually, in the trivial case $r=1$,
the moments $\Gamma(r,q)$ are always equal to $1$ for all kinds of
conservative multifractal measures and all orders $q$. We calculated
the values of $\min_r\psi(r,q)$ for different $q$ and $m$. The
results are shown in Fig.~\ref{Fig7}. We find that $\min_r\psi(r,q)$
approaches closer and closer to $1$ with the increase of $|q|$ or
the decrease of $m$. The distance of the minima $\min_r\psi(r,q)$ to
$1$ is exponential as a function of $q$ for fixed $m$. The slopes of
the straight lines fitted to the data for each $m$ in
Fig.~\ref{Fig7} are approximately evenly spaced for small $m$. This
verifies the analytic results that $\psi(\lambda^{-i}) = 1$.

\begin{figure}
\begin{center}
\includegraphics[width=8cm]{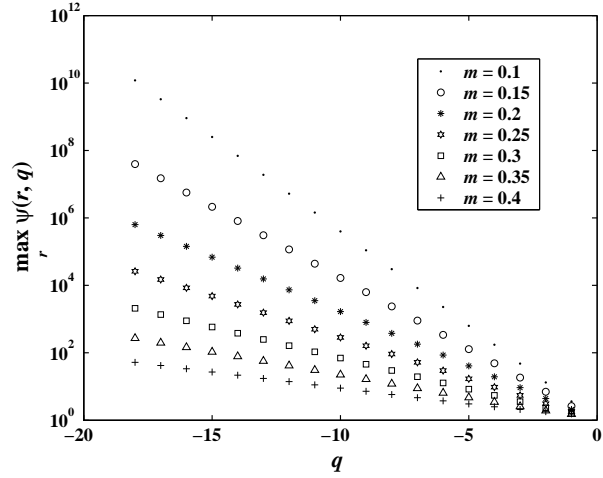}
\end{center}
\caption{\label{Fig8}The exponential dependence of $\max_r\psi(r,q)$
as a function of $q$ for different $m$. The slopes of the straight
lines in this log-linear plot are proportional to $\ln m$.}
\end{figure}

Figure \ref{Fig8} shows the local maxima $\max_r\psi(r,q)$ versus
$q$ for different $m$. Note that moments of larger $m$ contain a
large noise amplitude making the numerical determination of local
maxima rather difficult. We find that $\max_r\psi(r,q)$ increases
exponentially with $|q|$. This result still holds for other
lattice multifractal measures. The slopes of the straight lines
fitted to the data for each $m$ in the log-linear plot of
Fig.~\ref{Fig8} are proportional to $\ln m$.

We define the amplitude of log-periodic oscillations as
\be
h(q) =
\max_r\psi(r,q) - \min_r\psi(r,q).
\label{Eq:hq}
\ee
It is easy to
see that $h(q)$ has a behavior similar to that of $\max_r\psi(r,q)$.
It follows the important result that the amplitude of the
log-periodic oscillations of lattice multinomial measures increases
with $|q|$. A similar result is known for (lattice) fractals
\cite{Smith-Fournier-Spiegel-1986-PLA}. This may be related
in a qualitative way to the observation that large log-periodic
oscillations are found in DLA and in rupture processes
compared to very small log-periodic amplitudes in spin models
defined on hierarchical geometries. Indeed, fractal growth and
rupture processes are controlled by local field / stress concentration,
i.e., by the high-order moments of the field involved in the
construction of the patterns.

\section{\label{s1:DSItest}A novel approach for extracting
log-frequency}

A natural question arises in practice concerning the determination
of the log-frequency and thus the preferred scaling ratio in the
detection of log-periodicity in fractals or multifractals. Up to
now, five different methods have attempted to improve the detection
of log-periodicity \cite{Zhou-Sornette-2002-PD}. The common part of
four non-parametric methods among the five methods
is the Lomb periodogram analysis. The
Lomb analysis of $\psi(r,q)$ in Fig.~\ref{Fig6} gives a very
significant peak at log-frequency $f_0 = 1/\ln\lambda$. Here, we
study more complicated measures for which the log-periodic
oscillations are not visible to the bare eyes but can still be
detected by spectral analysis. We develop in this section a novel
method for the detection of log-frequencies, which is expected to be
powerful also more generally in the search of log-periodicity in
real systems.

We construct a triadic multifractal measure which has three
partitions in each generation with scaling ratios $\lambda_1 = 2$
and $\lambda_2 = \lambda_3 = 4$ and measure multipliers $m_1 = m_3
= 0.5$ and $m_2 = 0$. The measure is supported by the generalized
Cantor set. Thus the preferred scaling ratio is $\lambda = 2$ and
the fundamental log-periodic frequency is $f_0 = 1/\ln2$. The
analytic value of $\tau(q)$ is also given by (\ref{Eq:tau}) with
$m = 0.5$ and $\lambda = 2$. We use $n=11$ iterations for
the construction of the measure. Again, the counting box size is
evenly sampled in the logarithm of the scale. The moments $\Gamma(r,q)$ of the
measure were thus calculated for different orders $q$ based on the
box-counting method.

\begin{figure}
\begin{center}
\includegraphics[width=8cm]{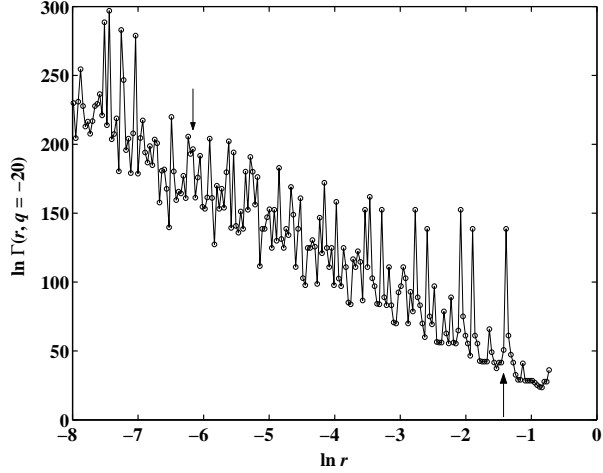}
\end{center}
\caption{\label{Fig9}The dependence of $\ln\Gamma(r,q)$ with respect
to $\ln r$ for $q=-20$. Although the plot is very noisy, a linear
least square fitting gives a slope close to the analytic exponent
$\tau(q)$. No unambiguous (log-)periodic oscillations are visible.}
\end{figure}

We present in Fig.~\ref{Fig9} a typical plot of $\ln\Gamma(r,q)$
versus $\ln r$ for $q<0$. Although the plot is very noisy, a
linear least square fitting gives a slope close to the analytic
exponent $\tau(q)$. No unambiguous simple (log-)periodic
oscillations are visible. Since $\ln r_c = -n\ln\lambda_1 =
-7.62$, we restricted the interval of study to $-6.17 < \ln r$,
where the lower bound is indicated by the left downward arrow in
Fig.~\ref{Fig9}. Indeed, a behavior different from large $r$ can
be observed for small $r$. We also introduced an outer cutoff in
the large $r$ range as indicated by the right upward arrow in
Fig.~\ref{Fig9}. Our analysis for the detection of log-periodicity was
thus carried out in the interval confined within the two arrows,
containing a total of $N=131$ data points. Note that a different
but reasonable choice of this interval doesn't impact on the
results.

\begin{figure}[htb]
\begin{center}
\includegraphics[width=8cm]{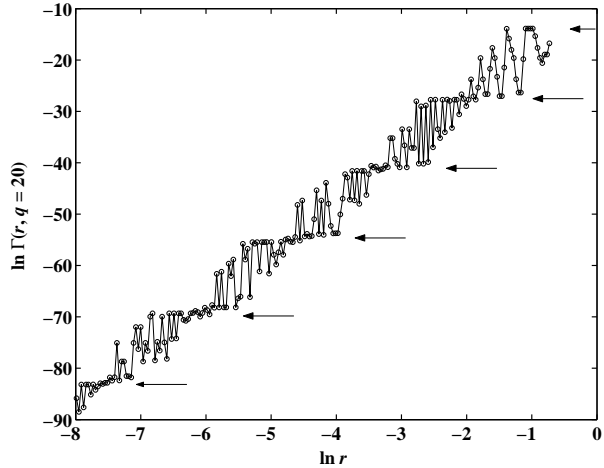}
\end{center}
\caption{\label{Fig10}The dependence of $\ln\Gamma(r,q)$ with
respect to $\ln r$ for $q=20$. The arrows indicate a periodic
pattern in $\ln r$. The period is $2\ln\lambda$ which is interpreted
as a subharmonic of the fundamental log-frequency.}
\end{figure}

Fig.~\ref{Fig10} shows a typical plot of $\ln \Gamma(r,q)$ versus
$\ln r$ for $q>1$.  A linear least square fit of the points recovers
a numerical estimate of $\tau(q)$ which is close to the theoretical
value. The plot also exhibits many parallel patterns
(rectangle-like) which are arranged periodically in $\ln r$, as
indicated by the arrows. We find that the period is approximately
$2\ln\lambda$, which is interpreted as the frequency of a
subharmonic to the fundamental log-frequency. A similar subharmonic
log-periodic frequency was also reportedly detected in the growing
diffusion-limited aggregation clusters
\cite{Sornette-Johansen-Arneodo-Muzy-Saleur-1996-PRL}. However, the
Lomb periodogram of the detrended $\ln \Gamma(r,q)$ in
Fig.~\ref{Fig10} doesn't provide convincing signal of such $f_0/2$
subharmonic (not shown). We nevertheless observed in the Lomb
periodogram two outstanding peaks at log-frequencies $f = 11.54$ and
$f = 15.87$, although this observation does not ensure the
statistical significance of log-periodicity. Lower order moments
provide more significant Lomb periodogram when $q > 0$, which means
that the noise was amplified rapidly for large $q$'s. On the other
hand, the impact of $q$ on the noise is less important when $q < 0$.
This noise amplification effect of larger $q$ suggests to
investigate lower-order moments to detect log-periodicity.

\begin{figure}
\begin{center}
\includegraphics[width=8cm]{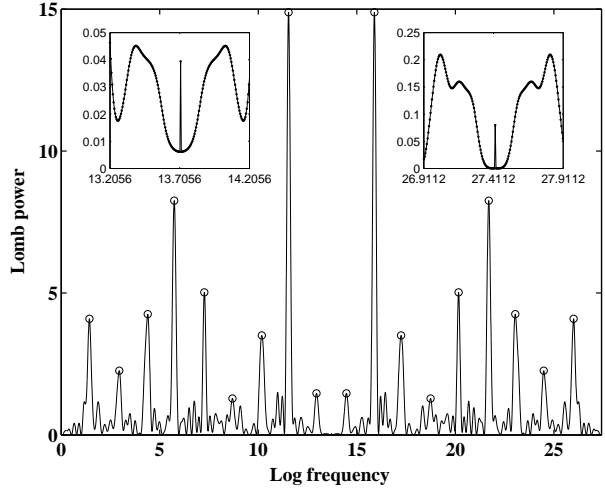}
\end{center}
\caption{\label{Fig11}The Lomb periodogram of the residuals of the
$q=-3$ order moment. The left inset shows the local details around
$f = 9.5/\ln \lambda$, which is the symmetric axis of the
periodogram. The right inset shows the local details around $f =
19/\ln\lambda$, beyond which the Lomb periodogram duplicates itself
periodically. The Lomb peaks indicated by open circles are
approximately evenly spaced.}
\end{figure}

We detrended the moment $\ln\Gamma(r,q=-3)$ to remove the
power-law scaling. Here the scaling exponent is estimated directly
from the data points and the resultant residuals have zero mean.
This method was used extensively in practice (this method is
different from that used in Sec.~\ref{s2:ResIII} and
\ref{s2:ResI}). Analyzing these residuals by the Lomb method gives
a Lomb periodogram shown in Fig.~\ref{Fig11}. It is interesting to
point out that the Lomb periodogram is periodic of period $19/\ln
\lambda = 27.4112$. The local detail around $f = 27.4112$ is shown
in the right inset of Fig.~\ref{Fig11}. We see a clear burst
exactly at $f = 27.4112$ causing the discontinuity at this point.
It is necessary to stress that this burst itself is not a Lomb
peak. It is thus sufficient to analyze only the $f<27.4112$ part
of the periodogram as shown in Fig.~\ref{Fig11}. In addition, the
Lomb periodogram in Fig.~\ref{Fig11} is apparently symmetric with
respect to $f = 9.5/\ln\lambda = 13.7056$. The left inset shows
the local detail around $f = 13.7056$, which is a discontinuous
burst as well. This is the reason why the right inset is symmetric
with respect to $f = 27.4112$.

In Fig.~\ref{Fig11}, the two highest peaks are located at $f =
11.54$ and $f = 15.87$. This is what we can see in the Lomb
periodogram for $q>1$ as we alluded above. The claim that the
fundamental logarithm frequency is $f_0 = 11.54$ or $f_0 = 15.87$,
as done in many other practical systems, has a false-alarm
probability of $0.006\%$ assuming that the noise is independent and
Gaussian
\cite{Horne-Baliunas-1986-ApJ,Press-Teukolsky-Vetterling-Flannery-1996}.
The existence of two peaks of the same height and the symmetric
structure of the Lomb periodogram reinforce further the statistical
significance of the existence of a genuine log-periodicity.

\begin{figure}
\begin{center}
\includegraphics[width=8cm]{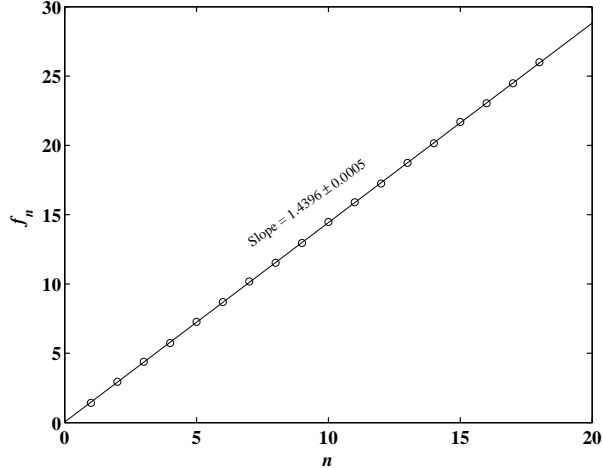}
\end{center}
\caption{\label{Fig12}The linear dependence of log-frequencies $f_n$
of the harmonics of the fundamental frequency $f_0$ as a function of
their order $n$. The slope of the fitted straight line is $1.4396\pm
0.0005$ providing an estimate of $f_0$ in excellent agreement with
the theoretic value of $f_0 = 1/\ln 2$. The preferred scaling ratio
is thus $\lambda = 2.0030 \pm 0.0005$ having only a $0.3\%$
deviation from the true value of $\lambda = 2$.}
\end{figure}

An additional important evidence is offered by expanding $\psi(r,q)$
in the Fourier space, which identifies a discrete set of
log-frequencies $f_n$ with amplitudes $A_n$. Here, we find that the
amplitudes $A_n$ of different frequencies $f_n$ are not restricted
to a presumed inequality that $A_1 > A_n$ with $n>1$ (we refer to
\cite{Zhou-Sornette-2002-IJMPC} for a detailed discussion). In the
case where several $A_n$ are comparable, as shown in
Fig.~\ref{Fig11}, we have to study the harmonics of the fundamental
frequency. There are indeed many slim peaks in Fig.~\ref{Fig11} that
are approximately evenly spaced. We indicated these $18$ peaks with
open circles, whose corresponding log-frequencies are denoted as
$f_n$. This recognition process is objective, since all except the
6th and 13th peaks are higher than the remaining peaks. It is quite
reasonable to interpret $f_n$ as the $n$-th harmonic of the
fundamental log-frequency. A convenient estimate of $f_0$ is the
mean of the differences of consecutive frequencies. This gives $f_0
= \langle f_n - f_{n-1} \rangle = 1.4457 \pm 0.0671$. In addition,
we can expect that $f_n \approx n f_0$. Thus, we can take the slope
of the line fitted to points $(n,f_n)$ as another estimate of $f_0$.
Figure \ref{Fig12} plots $f_n$ as a function of $n$, which exhibits
an almost perfect linear relationship. The slope of the fitted line
is $f_0 = 1.4396 \pm 0.0005$. The estimates of $f_0$ from the two
approaches are comparable to each other and both in excellent
agreement with the theoretic value of $f_0 = 1.4427$. However, the
linear regression method of Figure \ref{Fig12} is obviously much
more stable. The preferred scaling ratio is thus calculated to be
$\lambda = e^{1/f} = 2.0030$ with a standard deviation $\sigma_\lambda = \sigma_f
e^{1/f}/ f^2 = 0.0005$.

In real systems, it is possible that the Lomb periodogram is much
more noisy than found in Fig.~\ref{Fig11}. In that case, we may only
determine a few but not all harmonics, so that the successively
detected harmonics are not consecutive. In this case, taking the
differences between successively detected harmonics will not give a
constant value equal to the fundamental log-frequency. This
certainly increases the difficulty of the detection of
log-periodicity and lowers the significance of the extracted
log-frequency. One should thus estimate subjectively the possible
order of the detected harmonics and reconstruct the best match
according to a linear fit such as in Figure \ref{Fig12} but with
holes for the missing harmonics. Notwithstanding its difficulties,
this method was successfully applied to the search of log-periodic
in the energy dissipation of three dimensional fully developed
turbulence \cite{Zhou-Sornette-2002-PD}.

\section{\label{s1:concl}Concluding remarks}

We have reviewed and generalized in this paper the basic discrete
scale invariance equations concerning the moments $\Gamma(r,q)$ of
fractals, multifractal measures and joint multifractal measures
based on the presence of (statistical) self-similarity in these
objects. There exist a set $\mathcal{T}$ of complex dimensions for
fractals and complex exponents for multifractals, corresponding to
the set of complex solutions of the corresponding DSI equation (DSI:
discrete scale invariance). The
set $\mathcal{T}$ lies in a strip $[\tau^R(q),
\tau_0(q)]\times{\mathcal{R}}$ in the complex plane and is symmetric
with respect to the real axis, where $\tau^R(q) \in \mathcal{T}$ is
the only exponent on the real axis.

In the lattice case, the complex exponents are located evenly with
spacing $\omega$ (where $\omega$ is the fundamental angular
log-frequency) on finitely many vertical lines in the complex plane.
Corresponding to the complex exponents on the vertical line $\tau(q)
= \tau^R(q)$, there is a log-periodic expression of $\Gamma(r,q)$
with a fundamental log-frequency $f = 1/\ln\lambda$, where $\lambda$
is the preferred scaling ratio. We have verified the log-periodicity
in multifractals numerically and proposed a novel precise approach
for the extraction of $f$. Our simulations also show that the
signal-to-noise ratio of the moments increases with $|q|$ when $q<0$
or $q>0$. Since larger signal-to-noise ratio corresponds to more
significant Lomb peak, we should investigate lower-order moments to
extract log-frequencies, which is independent of the fact that
higher-order moments have larger amplitudes of oscillations.

In the non-lattice case, the complex exponents are irregularly
located in the strip
and $\tau^R(q)$ is the unique complex exponent in the
vertical line $\tau(q) = \tau^R(q)$. There are thus no preferred
scaling ratio and no exact log-periodicity. However, we can construct a
sequence of lattice multifractals whose preferred scaling ratios
converge to $1$ from above to approximate the non-lattice
multifractal so that the set of these lattice multifractals
converge to the set of complex exponents of the non-lattice
multifractal. In the Appendix, we propose an explicit algorithm for the
construction of the lattice sequence.

Therefore, we can recognize lattice fractals and/or multifractals
in real systems from the evidence of the presence of log-periodicity. We can obtain a
preferred scaling ratio $\lambda$ based on spectral analysis. This
search not only sheds new light towards a better understanding of
the system under study but also offers a constraint on the modelling
of the system. Of course, $\lambda$ does not contain a detailed
information on all the scaling ratios $\lambda_i$. It should be noted that the search for
log-periodicity has been carried out already in several real systems, such
as in the growing diffusion-limited aggregation (DLA) clusters
\cite{Sornette-Johansen-Arneodo-Muzy-Saleur-1996-PRL}, in two
dimensional free decaying turbulence
\cite{Johansen-Sornette-Hansen-2000-PD} and in three dimensional
fully developed turbulence
\cite{Sornette-1998,Zhou-Sornette-2002-PD}. It is interesting that
the preferred scaling ratio is $\lambda \approx 2$ for the cases of
DLA and 3D turbulence and it is possible that the log-frequency $f_0
= 4 \sim 5$ in the two dimensional free decaying turbulence case
\cite{Johansen-Sornette-Hansen-2000-PD} turns out to be a harmonic
of $f_0 = 1/\ln 2$. However, we have to stress that the existence of
a preferred scaling ratio $\lambda$ does not imply that there is
only one scaling ratio equal to the measured $\lambda$. For
instance, $\lambda = 2$ found in 3D turbulence does not mean that
the $p$-model \cite{Meneveau-Sreenivasan-1987-PRL} is the correct
and only one model. We shall further clarify this point by taking
the growing DLA clusters as an example as follows.

There are many models proposed to explain the fractal growth of DLA
clusters, such as the hierarchical ghost model
\cite{Ball-Witten-1984-JSP,Witten-Cates-1986-Science} and the
Fibonacci model
\cite{Arneodo-Argoul-Bracy-Muzy-1992-PRL,Arneodo-Argoul-Muzy-Tabard-1992-PLA,Arneodo-Argoul-Muzy-Tabard-Bacry-1993-Fractals}.
In the hierarchical ghost model, pairs of single particles collide
and aggregate to form dimers. Then pairs of dimers collide and
aggregate to form four-particle clusters. This hierarchical growth
process goes on and on to form eventually a large DLA cluster.
Defining $b(N)$ as the average number of bonds along the connecting
pathway between two particles on an $N$-particle cluster, we obtain
the DSI equation \cite{Ball-Witten-1984-JSP} \be b(2N) =
\frac{3}{2}b(N). \label{Eq:DLA1} \ee Since the average radius $R$ of
a $N$-particle cluster obeys $R \sim \sqrt{b(N)}$ and $N \sim R^D$,
we have $D = 2\ln2/\ln(3/2)$ and the set of complex exponents with
respect to (\ref{Eq:DLA1}) is \be {\mathcal{T}} = \{2/D + 2\pi n i /
\ln 2,~n\in{\mathcal{Z}}\}. \label{Eq:DLA1T} \ee It is clear that
$\lambda = 2$ and the set of complex exponents is on a vertical line
with abscissa of $2/D$ in the complex plane. On the other hand, by
applying the results of the wavelet transform modulus maxima (WTMM)
representation to the DLA azimuthal Cantor set, it follows that the
number $N(a)$ of maxima of the WTMM skeleton at scale $a$ satisfies
the Fibonacci rule
\cite{Arneodo-Argoul-Bracy-Muzy-1992-PRL,Arneodo-Argoul-Muzy-Tabard-1992-PLA,Arneodo-Argoul-Muzy-Tabard-Bacry-1993-Fractals}:
\be N(a) = N(\lambda a) + N (\lambda^2 a). \label{Eq:DLA2} \ee It is
easy to show that the set of complex dimensions is
\cite{Sornette-Johansen-Arneodo-Muzy-Saleur-1996-PRL,Lapitus-Frankenhuysen-2000}
\be {\mathcal{D}} = \{1+\frac{\ln\phi}{\ln\lambda} + i\frac{2\pi n
}{\ln\lambda}\} \cup \{1-\frac{\ln\phi}{\ln\lambda} +
i\frac{(2n+1)\pi}{\ln\lambda}\}, \label{Eq:DLA2D} \ee where
$n\in{\mathcal{Z}}$ and $\phi = (\sqrt{5}+1)/2$ is the golden mean.
We see that there are two vertical lines in the complex plane in
this model. Numerical simulations
\cite{Sornette-Johansen-Arneodo-Muzy-Saleur-1996-PRL} lead to the
identification of two log-frequencies $f_0=1/\ln\lambda$ and $f_0/2$
in excellent agreement with the prediction of the Fibonacci model.
The preferred scaling ratio was estimated numerically to be $\lambda
= 2.2$. It shows that the Fibonacci model with two scaling ratios
$\lambda$ and $\lambda^2$ provides a better description than the
hierarchical ghost model in the sense of discrete scale invariance.

%

\pagebreak

\appendix\section{General framework of discrete scale
invariance in fractals and multifractals} \label{s1:framework}

In this Appendix, we revisit the equations governing the concepts of
discrete scale invariance (DSI) and the associated complex exponents
in fractals and multifractal measures, and generalize them to joint
multifractal measures. We shall show that the set of complex
exponents lie symmetrically with respect to the real axis in a
vertical strip in the complex plane. Furthermore, in the lattice
case, the set of complex exponents lie on finitely many vertical
lines periodically and the DSI leads to the log-periodic corrections
to scaling implying a preferred scaling ratio that is proved to be
independent of the order of the moments. In the non-lattice case,
there are no visible log-periodicity, {\em{i.e.}}, no preferred
scaling ratio since the set of complex exponents are spread
irregularly within the complex plane. A non-lattice multifractal can
be approximated by a sequence of lattice multifractals so that the
sets of complex exponents of the lattice sequence converge to the
set of complex exponents of the non-lattice one. An algorithm for
the construction of the lattice sequence is proposed explicitly.

\subsection[A]{Revisiting DSI in fractals}
\label{s2:FDSI}

Smith et al \cite{Smith-Fournier-Spiegel-1986-PLA} studied three
Cantor sets with the initiators being 101 (the classic Cantor set),
101001001, and 101010001. They followed the Grassberger-Procaccia (G-P) approach
\cite{Grassberger-Procaccia-1983-PD} and investigated the periodic
oscillations in the log-log plot of the correlation integral
$C(\ell)$ with respect to scale $\ell$. In this section, we
revisited DSI in the classic Cantor set using an alternative
approach \cite{Bessis-Geronimo-Moussa-1983-JPL}.

In order to generalize to other self-similar fractal sets, consider
a fractal set ${\mathcal{F}}$, which is the union of $n$ disjoint
parts ${\mathcal F}_i$ , namely ${\mathcal F} = \cup_{i=1}^n
{\mathcal F}_i$ with ${\mathcal F}_i \cap {\mathcal F}_j = \Phi$ for
$i\neq j$. Let $N({\mathcal{F}};r)$ be the number of boxes covering
${\mathcal F}$ with scale $r$. We assume that the magnification of
${\mathcal F}_i$ by a factor $\lambda_i$ results in ${\mathcal F}$.
This leads to \be N({\mathcal{F}}_i; r) = N({\mathcal{F}}; \lambda_i
r). \label{Eq:FSS} \ee Note that the reciprocal of the amplification
factor, $1/\lambda_i$, is called the scale multiplier. Since \be
N({\mathcal{F}}; r) = \sum_{i=1}^n N({\mathcal{F}}_i; r),
\label{Eq:FSum} \ee we obtain \be N({\mathcal{F}}; r) = \sum_{i=1}^n
N({\mathcal{F}}; \lambda_i r), \label{Eq:FDSI} \ee Equation
(\ref{Eq:FDSI}) has a solution if and only if there exist integers
$k_i$ and $\lambda$ satisfying \be \lambda_i = \lambda^{k_i},
\label{Eq:LattCond} \ee which is the generalization of the Fibonacci
rule of the number of maxima of the WTMM skeleton at a given scale
in diffusion-limited aggregation
\cite{Sornette-Johansen-Arneodo-Muzy-Saleur-1996-PRL} and
corresponds to the so-called lattice case
\cite{Lapitus-Frankenhuysen-2000}. The solution can be expressed as
   \be N({\mathcal{F}};r) = r^{-D}
\psi (r). \label{Eq:FSol} \ee where $\psi(r)=\psi(\lambda r)$ and
$D$ is the fractal dimension determined by the real solution of
equation \be \sum_{i=1}^n \lambda_i^{-D} = 1. \label{Eq:D} \ee We
see that $N({\mathcal{F}};r)$ exhibits log-periodic oscillations
with respect to $r$. Since $N({\mathcal{F}},1)=1$, it follows that
$\psi(\lambda^{-\ell})=1$, where $\ell \in \overline{Z^-}$. For $k_i
= 1$, we recover Eq.~(\ref{Eq:O}).

\subsection{DSI in multinomial measures}
\label{s2:MfDSI}

\subsubsection{General equation of DSI in multinomial measure}
\label{s3:Mfeqn}

We construct a self-similar multinomial measure $\mu$ supported by a
set ${\mathcal{F}}$ by a repeated $n$-based multiplicative cascade,
where $\mathcal F$ can be an Euclidean set or a fractal set of
dimension $\dim{\mathcal{F}}$. In the first step, the support
${\mathcal{F}}$ carrying some measure $\mu$ is partitioned into $n$
small pieces ${\mathcal{F}}_i$ with $i=1,\cdots,n$ based on some
given set of scale amplification factors \be \lambda_i =
\parallel {\mathcal{F}} \parallel / \parallel {\mathcal{F}}_i \parallel,
\label{Eq:Mflambda} \ee where the measure is redistributed on
${\mathcal{F}}_i$ according to a set of given multipliers: \be m_i =
\mu({\mathcal{F}}_i; r/\lambda_i) / \mu({\mathcal{F}}; r).
\label{Eq:Mfm} \ee In the next step, each piece ${\mathcal{F}}_i$ is
further partitioned into $n$ smaller pieces according to the same
amplification factors $\lambda_i$ and its carrying measure is
redistributed again with the same multipliers $m_i$. This procedure
continues {\it{ad}} infinity. So far, we have constructed a
deterministic self-similar multinomial measure. Define a $q$-order
moment function \be \Gamma({\mathcal{F}};r,q) = \sum_r
\mu^q({\mathcal{F}}; r), \label{Eq:MfMom} \ee where
$\mu({\mathcal{F}}; r)$ is the measure in an arbitrary box of scale
$r$ and $\sum_r$ represents the sum over all boxes which result from
partitioning ${\mathcal{F}}$ with scale $r$. When we use boxes of
size $r$ to cover ${\mathcal{F}}$, the minimum number of disjoint
boxes is $[\parallel{\mathcal{F}}\parallel/r]$. One finds easily
that there might exist a box covering a part of ${\mathcal{F}}$
whose size is less than $r$. In this case, we can arrange the
$[\parallel{\mathcal{F}}\parallel/r]$ boxes in a row (like a string)
and cover ${\mathcal{F}}$ in infinitely different ways by
translating the boxes string along ${\mathcal{F}}$, thus introducing
a random ``phase.'' This phase is one source of noise in the
box-counting method. When $r$ is small enough, the noise is reduced
greatly so that the moments corresponding to different phases are
approximately identical. In other words, the moment function is not
affected by the partition procedure and we have \be
\Gamma({\mathcal{F}};r,q) = \sum_{i=1}^n
\Gamma({\mathcal{F}}_i;r,q)~. \label{Eq:MfSum} \ee Substitution of
Eq.~(\ref{Eq:Mfm}) in Eq.~(\ref{Eq:MfMom}) leads to \be
\Gamma({\mathcal{F}}_i;r,q) = m_i^q \Gamma({\mathcal{F}};\lambda_i
r,q), \label{Eq:MfSS} \ee which shows the self-similarity of the
moments. Combination of Eqs.~(\ref{Eq:MfSum}) and (\ref{Eq:MfSS})
gives \be \Gamma({\mathcal{F}};r,q) = \sum_{i=1}^n
\Gamma({\mathcal{F}};\lambda_ir,q)m_i^q. \label{Eq:MfDSI} \ee This
is the basic DSI equation of the deterministic self-similar
multinomial measures. For statistically self-similar measures
\cite{Falconer-1994-JTP}, Eq.~(\ref{Eq:MfDSI}) can be revised as \be
\Gamma({\mathcal{F}};r,q) = \sum_{j=1}^k \sum_{i=1}^n
\Gamma({\mathcal{F}};\lambda_{i,j}r,q)m_{i,j}^q p_j,
\label{Eq:MfDSIRnd} \ee where $p_j$ is the probability of choosing
the $j$-th rule out of the $k$ rules and $\lambda_{i,j}$ and
$m_{i,j}$ are accordingly the scaling ratios and measure multipliers
\cite{Zhou-Yu-2002}.

\subsubsection{Solution of the DSI equation for lattice multifractals}
\label{s3:Mfsol}

In the case of $q=0$, the moment function equals to the number of
partitioned boxes $\Gamma({\mathcal{F}};r,0) = N({\mathcal{F}}; r)$,
which recovers Eq.~(\ref{Eq:FDSI}).

If $\lambda_1 = \cdots = \lambda_n \equiv \lambda$, it follows from
Eq.~(\ref{Eq:MfDSI}) that \be \Gamma({\mathcal{F}};r,q) =
\Gamma({\mathcal{F}};\lambda r,q) \sum_{i=1}^n m_i^q.
\label{Eq:MfDSI2} \ee In the formalism of multifractals, one expects
that \be \Gamma({\mathcal{F}};r,q) \propto r^{\tau(q)},
\label{Eq:Mftau} \ee where the scaling exponent of the moment is in
the form \be \tau(q) = -\ln \sum_{i=1}^n m_i^q / \ln \lambda.
\label{Eq:MfRtau} \ee In general, the solution to
Eq.~(\ref{Eq:MfDSI2}) is \be \Gamma({\mathcal{F}};r,q) = r^{\tau(q)}
\psi(r), \label{Eq:MfSol} \ee where $\psi(r)=\psi(\lambda r)$ and
$\tau(q)$ is defined in Eq.~(\ref{Eq:MfRtau}). Since
$\Gamma({\mathcal{F}};1,q)=1$, we have $\psi(\lambda^{-\ell}) = 1$,
where $\ell\in\overline{Z^-}$.

Now, consider the case of lattice multifractals with the scale
multipliers $\lambda_i$ satisfying condition (\ref{Eq:LattCond}). It
follows from (\ref{Eq:MfDSI}) that \be \Gamma({\mathcal{F}};r,q) =
\sum_{i=1}^n \Gamma({\mathcal{F}};\lambda^{k_i}r,q)m_i^q.
\label{Eq:MfDSILatt} \ee It is easy to verify that
Eq.~(\ref{Eq:MfSol}) is also the general solution to
Eq.~(\ref{Eq:MfDSILatt}), where $\tau(q)$ is the unique real
solution of \be \sum_{i=1}^n \lambda^{k_i\tau(q)} m_i^q = 1,
\label{Eq:MftauLatt} \ee We emphasize that $\psi$ is related to $q$ but
independent of $\lambda$. Also, the general solution of
Eq.~(\ref{Eq:MfDSIRnd}) has the same form of (\ref{Eq:MfSol}) with
different ``$\tau(q)$-generation function'' in the lattice case.

\subsubsection{Oscillations of other characteristic functions
of multifractals} \label{s3:Mffuns}

It is natural to investigate the presence of oscillations for other functions
characterizing multifractals, for instance, the generalized dimensions
$D_q(r)$, the strengths of singularity $\alpha(r,q)$, and the
multifractal function $f(r,\alpha)$ or $\tilde{f} (r,q)$
\cite{Grassberger-1983-PLA,Hentschel-Procaccia-1983-PD,Frisch-Parisi-1985,Halsey-Jensen-Kadanoff-Procaccia-Shraiman-1986-PRA},
when one investigates scaling laws in some experimental data as well
as simulated data. Our analysis is performed in terms of the
variable $r$ since all these functions vary with $r$. As usual, we
define the exponent function of the moments as \be \tau(r,q) = d \ln
\Gamma(r,q) /d\ln r. \label{Eq:MfDeftaur} \ee In equation
(\ref{Eq:MfDeftaur}), we remove the symbol of the support
$\mathcal{F}$ and use $\tau(r,q)$ instead of $\tau(q)$ as is used in
\cite{Grassberger-1983-PLA,Hentschel-Procaccia-1983-PD}. Simple
algebraic calculations on equation (\ref{Eq:MfSol}) leads to \be
\tau(r,q) = \tau(q) + \Psi(r,q), \label{Eq:Mftaur} \ee where \be
\Psi(r,q) = \frac {r\psi'(r)}{\psi(r)}, \label{Eq:MfPsi} \ee
Similarly, we have \be D_q(r) \stackrel{\triangle}{=} \lim_{q' \to
q} \frac {\tau(r,q)}{ q'-1} = D_q + \lim_{q' \to q} \frac
{\Psi(r,q')}{q'-1}, \label{Eq:MfDqr} \ee \be \alpha(r,q)
\stackrel{\triangle}{=} \frac{\partial \tau(r,q)}{
\partial q} = \alpha(q)+ \frac {\partial \Psi(r,q) }{ \partial q},
\label{Eq:Mfalphar} \ee and
\begin{equation}
 \tilde{f}(r,q) \stackrel{\triangle}{=} q\alpha(r,q) - \tau(r,q) \tilde{f}(q) + q
 \frac {\partial \Psi(r,q)}{ \partial q} - \Psi(r,q). \label{Eq:Mffr}
\end{equation}
Hence, the four characteristic functions of multifractals,
$D_q(r,q)$, $\tau(r,q)$, $\alpha(r,q)$, and $\tilde{f}(r,q)$, are
all log-periodic in $r$ with the same period $\ln \lambda$, which is
independent of the order $q$. This result was verified in an
experimental situation by extracting the log-periodicity of the
moments of the energy dissipation in three-dimensional fully
developed turbulence \cite{Zhou-Sornette-2002-PD}. We shall return
to this issue in Sec.~\ref{s2:CE}.

\subsection{DSI in joint multifractal measures}
\label{s2:JDSI}

In complex systems, there may be more than one self-similar measure
embedded in the same sample space or phase space. For instance, in
heated turbulent flow, both the energy dissipation rate $\epsilon$
and scalars, say concentration $C$ or temperature $T$, are known to
be multifractal. Thus, it is interesting to consider the joint
multifractal measures. Suppose that $\mu_j$ with $j = 1, \cdots, l$
are $l$ self-similar measures embedded in $\mathcal{F}$. The joint
moments of order $q = \sum_{j=1}^l q_j$ of the $l$ measures are
given by \be J({\mathcal{F}}; r, q) = \sum_r \prod_{j=1}^l \mu_j
^{q_j}({\mathcal{F}}; r). \label{Eq:J} \ee The measure multipliers
of $\mu_j$ are defined by \be m_{i,j} =
\mu_j({\mathcal{F}};r/\lambda_i)/\mu_j({\mathcal{F}};r).
\label{Eq:Jmij} \ee Combining (\ref{Eq:J}) and (\ref{Eq:Jmij}), we
have \be J({\mathcal{F}}_i; r, q) = J({\mathcal{F}};\lambda_i r,
q)\prod_{j=1}^l m_{i,j}^{q_j}. \label{Eq:Jss} \ee It follows from
the additive rule of the joint moment \be J({\mathcal{F}}; r, q) =
\sum_{i=1}^nJ({\mathcal{F}}_i; r, q) \label{Eq:JJi} \ee that the DSI
equation is \be J({\mathcal{F}}; r, q) = \sum_{i=1}^n
J({\mathcal{F}}; \lambda_{i} r, q) \prod_{j=1}^l m_{i,j}^{q_j},
\label{Eq:JDSI} \ee for deterministic measures where $\lambda_{i,j}$
and $m_{i,j}$ are the $i$-th multipliers of $\mu_j$ or \be
J({\mathcal{F}}; r, q) = \sum_{k=1}^{n'} \sum_{i=1}^{n}
J({\mathcal{F}}; \lambda_{i,k} r, q) \prod_{j=1}^{l} m_{i,j,k}^{q_j}
p_k, \label{Eq:JDSIRnd} \ee for statistically self-similar measures
where $\lambda_{i,k}$ and $m_{i,j,k}$ are $i$-th multipliers of
$\mu_j$ for $k$-th rule with probability $p_k$.

Similarly, if $\mu_j$ satisfy the lattice condition so that
$\lambda_{i,k} = \lambda^{k_{i,k}}$, the general solution of
Eq.~(\ref{Eq:JDSIRnd}) is \be J({\mathcal{F}}; r, q) = r^{\tau(q_1,
\cdots, q_l)} \psi(r), \label{Eq:JSol} \ee where
$\psi(r)=\psi(\lambda r)$ and $\tau(q_1, \cdots, q_l)$ is the unique
real solution of the following equation: \be \sum_{k=1}^{n'}
\sum_{i=1}^n \lambda_{i,k}^{\tau(q_1, \cdots, q_l)} \prod_{j=1}^l
m_{i,j,k}^{q_j}p_k = 1. \label{Eq:Jtau} \ee Again, since
$J({\mathcal{F}};1,q)=1$, we have $\psi(\lambda^{-\ell}) = 1$, where
$\ell\in\overline{Z^-}$. The deterministic case Eq.~(\ref{Eq:JDSI})
can be viewed as a special case of Eq.~(\ref{Eq:JDSIRnd}) with
$n'=1$.

It is clear that the DSI equations of multifractal measures in
Sec.~\ref{s2:MfDSI} are special cases of those of the joint
multifractal measures with $l=1$. Moreover, their DSI equations are
the same in essence despite the possible difference in the
coefficients. We expect to confirm the relevance of these formulas
in real systems with hierarchical structures, especially in
turbulence.

\subsection{\label{s2:CE}Complex exponents}

The previous sections emphasized the log-periodic structure of
lattice fractals and multifractals. However, there is no
log-periodicity for the non-lattice case. Instead, we can adopt the
more general concept of complex exponents as was done in the former
literature
\cite{Makarov-1994-JMAA,Sornette-1998-PR,Lapitus-Frankenhuysen-2000}.
Indeed, log-periodicity is only a special case of complex exponents
in the lattice case, while complex exponents exist for both lattice
and non-lattice fractals and multifractals. The complex dimensions
of fractal strings were extensively studied from a mathematical
view point \cite{Makarov-1994-JMAA,Lapitus-Frankenhuysen-2000}. We note
that a fractal string is not identical to the original fractal but
to its complement. Hence, we propose a different approach, even for
the fractals discussed in the previous sections. In this section, we
study the structure of complex exponents in the complex plane and
then provide an explicit algorithm for the construction of a
sequence of lattice multifractals approximating a non-lattice
multifractal whose set of complex exponents is the limit set of the
sets of complex exponents of the lattice sequence.

\subsubsection{\label{s3:struct} The structure of complex
exponents}

In the previous sections, we assume that $\tau(q)$ is real, thus the
prefactor of the scaling relation $\Gamma(r,q) \propto r^{\tau(q)}$
is a function of $r$ in the lattice case, as expressed in
Eq.~(\ref{Eq:MfSol}). Expanding $\psi$ in the Fourier space
yields complex exponents \cite{Sornette-1998-PR}. Therefore, we can
assume that $\tau(q)$ is complex so that the prefactor of the
scaling relation is independent of $r$. In this sense, we have \be
\sum_{i=1}^n m_i^q \lambda_i^{\tau(q)} = 1.\label{Eq:CE} \ee A more
general formulae is Eq.~(\ref{Eq:Jtau}). The solutions
of Eq.~(\ref{Eq:CE}) in the complex plane are denoted as \be \tau(q)
= \tau_R(q) + i\tau_I(q), \label{Eq:tauRI} \ee where $i = \sqrt{-1}$
is the imaginary unit. We shall study this Eq.~(\ref{Eq:CE}) in the
sequel. The results are the same for the stochastic case.

For convenience, let $h(\tau) = \sum_{i=1}^n m_i^q \lambda_i^{\tau}
- 1$. Since $h(-\infty) = \infty$ and $h(\infty) = -1$, and because
$h(\tau)$ is continuous, real solutions to (\ref{Eq:CE}) exist.
Moreover, $h(\tau)$ is differentiable and we have $dh(\tau)/d\tau <
0$. Thus Eq.~(\ref{Eq:CE}) has a unique real solution, denoted as
$\tau^R(q)$ hereafter. The geometric properties of $\tau^R(q)$ are
well-known
\cite{Halsey-Jensen-Kadanoff-Procaccia-Shraiman-1986-PRA}. As for
the complex solutions, it is easy to show from the real value of
$h(\tau)$ that, if $\tau(q)$ is a solution to (\ref{Eq:CE}), then
$\overline{\tau}(q) = \tau_R(q) - i\tau_I(q)$ is also a solution.
Therefore the set of complex exponents is symmetric with respect to
the real axis (the complex exponents come in pairs of complex conjugates).

Let $\tau(q)$ be a solution of Eq.~(\ref{Eq:CE}). Then $\tau_R(q)
\ge \tau^R(q)$. Otherwise, by assuming that $\tau_R(q) < \tau^R(q)$,
we have $|\sum_{i=1}^n m_i^q \lambda_i^{\tau(q)}| \leq \sum_{i=1}^n
m_i^q \lambda_i^{\tau_R(q)} < \sum_{i=1}^n m_i^q
\lambda_i^{\tau^R(q)} = 1$. This means that the complex exponents
lie on the right of the vertical line $\tau_R(q) = \tau^R(q)$. In
the lattice case, $\tau(q) = \tau^R(q) + in\omega$ also belong to
the set of complex exponents, where $n\in \mathcal{Z}$ and $\omega =
2\pi/\ln\lambda$ is the fundamental angular log-frequency of lattice
fractals and multifractals. In the non-lattice case, $\tau(q) =
\tau^R(q)$ is the only complex exponent satisfying $\tau_R(q) =
\tau^R(q)$. Furthermore, there is a right boundary $\tau_0(q)$ of
the set of complex exponents, that is, $\tau_R(q) < \tau_0(q)$. (We
refer the interested reader to Sec. 2.5 of
\cite{Lapitus-Frankenhuysen-2000} for a rigorous proof for fractal
strings, which can be easily generalized to the present case.) Hence
the set of complex exponents lie in a strip $[\tau^R(q),\tau_0(q)]
\times {\mathcal{R}}$.

In the case of lattice multifractal, there exists a preferred
scaling ratio $\lambda$ so that equation (\ref{Eq:CE}) can be
expressed as a polynomial equation (\ref{Eq:MftauLatt}) with respect
to the unknown $\lambda^{\tau(q)}$ of degree $\max_{j} \{ k_j\}$.
The set of complex exponents is obtained easily from the set of
complex solutions of (\ref{Eq:MftauLatt}). Hence there are finite
complex numbers $\tau_j(q)$ such that the set of complex exponents
is given by \be {\mathcal{T}} = \{\tau_j(q) + in\omega,~n \in
{\mathcal{Z}},~ j = 1, 2, ..., \ell\}, \label{Eq:T} \ee where $\ell
\leq \max_{j} \{ k_j\}$. Thus the complex exponents lie on finitely
many vertical lines in the complex plane. The points on each
vertical line corresponding to a given $\tau_j(q)$ are distributed
evenly with spacing $\omega$.

Assume that $(\tau_j)_R \neq (\tau_k)_R$ for $j\ne k$ and $\tau_1(q)
= \tau^R(q)$. Then we have \be (\tau_j)_I = \omega/2,
\label{Eq:taujI} \ee where $j>1$. This result derives because the
points on each line are evenly located with spacing $\omega$ and
symmetric with respect to the real axis. The log-frequencies of the
moments are thus given by \be f_{j,n} = \frac{(\tau_j)_I +
n\omega}{2\pi} = \left\{ {\begin{array}{cc}
{n\omega}/{2\pi},& {\mathtt{if}}~ j = 1\\
(\frac{1}{2}+n) \omega/2\pi,& {\mathtt{if}}~ j > 1 \end{array}}
\right., \ee where $n\in {\mathcal{Z}}-\{0\}$ when $j=1$ and $n \in
{\mathcal{Z}}$ when $j>1$, $f_{1,1}$ is the fundamental
log-frequency and $f_{j>1,0}$ are subharmonics of $f_{1,1}$
\cite{Sornette-Johansen-Arneodo-Muzy-Saleur-1996-PRL}. Certainly, it
is also possible that there are at least two complex exponents with
indices $j$ and $k$ such that $(\tau_j)_R = (\tau_k)_R$. In this
case, the points on the line $\tau_R(q) = (\tau_k)_R$ are no
longer evenly spaced with period $\omega$. This possibility will be
clearer in the following when we approximate a non-lattice
multifractal with a sequence of lattice multifractals. It is clear
that $f_{j>1,0}$ might change for different orders $q$.
Nevertheless, $f_{1,1} = {\omega}/{2\pi}$ holds for any $q$. In
other words, the fundamental log-frequency of a lattice multifractal
is independent of the order $q$ \cite{Zhou-Sornette-2002-PD}.

In the case of non-lattice multifractals, the set of complex
exponents is not periodic. However, the necessary and sufficient
condition for the existence of log-periodicity is the existence of
periodicity in the plane of the complex exponents associated with a
preferred scaling ratio. In the present case, there is no such
preferred scaling ratio and hence no log-periodicity. The complex
exponents of a non-lattice multifractal can be approximated by the
complex exponents of a sequence of lattice multifractals with larger
and larger $\omega$. The approximation sequence of lattice
multifractals can be constructed explicitly, as in the case of
fractal strings \cite{Lapitus-Frankenhuysen-2000}. This construction
process is described in the following Sec.~\ref{s3:approx}.

\subsubsection{\label{s3:approx} Approximating non-lattice
multifractal by a sequence of lattice multifractals}

Let us denote the scaling ratios $\Lambda$ of the non-lattice
multifractal as $\lambda_j$ where $j=1,\cdots,n$. Given any real
number $\lambda_0>1$, we have a set of numbers $\Theta = \{\theta_j
= \ln\lambda_j/\ln\lambda_0: j=1,\cdots,n\}$, defining the
coordinates of a point in the $n$ dimensional space
${\mathcal{R}}^n$. Since the multifractal is non-lattice, there is
at least one $j$ such that $\theta_j$ is an irrational number for
any possible choice of $\lambda_0$. It is known that there exists a
sequence of points $\Theta^{(k)} = \{\theta^{(k)}_j: j=1,\cdots,n\}
\in {\mathcal{R}}^n$ which converge to $\Theta$, where
$\theta^{(k)}_j$ are rational numbers. Obviously, the sequence of
rational numbers $\theta^{(k)}_j$ converge to $\theta_j$ for every
$j=1,\cdots,n$. If some $\theta_j$ is rational, we stipulate
trivially that $\theta^{(k)}_j = \theta_j$ for all $k$. Posing
$\lambda^{(k)}_j = \lambda_0^{\theta^{(k)}_j}$, we obtain a sequence
of lattice multifractals $\Lambda^{(k)}$ with scaling ratios
$\{\lambda^{(k)}_j: j=1,\cdots,n\}$ converging to the non-lattice
multifractal. The set of complex exponents ${\mathcal{T}}^{(k)}$ of
$\Lambda^{(k)}$ converges to the set of complex exponents
${\mathcal{T}}$ of $\Lambda$. In practice, we can take a lattice
multifractal $\Lambda^{(k)}$ with $k$ large enough as an
approximation of the non-lattice multifractal $\Lambda$.

Let us write $\theta^{(k)}_j = s^{(k)}_j/t^{(k)}_j$ where
$s^{(k)}_j$ and $t^{(k)}_j$ are relatively prime. Let $t^{(k)}$ be
the minimal common divisor of $\{t^{(k)}_j: j = 1,\cdots,n \}$. Then
$\lambda^{(k)} = \lambda_0^{1/t^{(k)}}$ is the preferred scaling
ratio of $\Lambda^{(k)}$. We thus recover the polynomial equation
(\ref{Eq:MftauLatt}): \be \sum_{j=1}^n m_j^q
\left[\lambda_0^{\tau(q)/t^{(k)}}
\right]^{s^{(k)}_jt^{(k)}/t^{(k)}_j} = 1, \label{eq:CE2} \ee where
${s^{(k)}_jt^{(k)}/t^{(k)}_j}$ are integers. We can choose a
sequence ensuring that $\theta^{(k)}_j$ is monotonous with respect
to $k$. It follows that $t^{(k)}$ increases with $k$. Therefore, the
fundamental log-frequencies $f^{(k)} = 1/\ln \lambda^{(k)}$ of
$\Lambda^{(k)}$ increase when converging to $\Lambda$. In other
words, there is no finite limit of $f^{(k)}$ when $k$ tends to
$\infty$ and the fundamental log-frequency of a non-lattice
multifractal is thus not defined. Practically, when one investigates
the Lomb periodogram of the residuals of a $q$-order moment of a
non-lattice multifractal, the log-frequency corresponding to the
highest Lomb peak is indefinite and increases when the sampling of
data points of the moment becomes denser (so as to investigate more
scales $r$) \cite{Zhou-Sornette-2002-PD}. On the other hand, the
distance of separation between the complex exponents on each
vertical line becomes larger and larger. This is not surprising
since the degree $\max_{j} \{{s^{(k)}_jt^{(k)}/t^{(k)}_j}\}$ of the
polynomial equation (\ref{eq:CE2}) increases with $k$ implying more
vertical lines, some of which may overlap.

\bibliography{E:/papers/Auxiliary/bibliography}

\end{document}